\documentclass{PoS}   %%JOURNAL%%

\title{Neutrino-driven wind and wind termination shock in supernova cores}

\ShortTitle{Neutrino-driven wind in supernova cores}
\author{\speaker{Almudena Arcones},  Leonhard Scheck,  Hans-Thomas Janka\\
  E-mail: \email{arcones@mpa-garching.mpg.de}\\
  Max-Planck-Institut f\"ur Astrophysik\\
  Karl-Schwarzschild-Stra{\ss}e 1\\
  D-85741 Garching, Germany}

\abstract{ The neutrino-driven wind from a nascent neutron star at the center
  of a supernova expands into the earlier ejecta of the explosion. Upon
  collision with this slower matter the wind material is decelerated in a wind
  termination shock.  By means of hydrodynamic simulations in spherical
  symmetry we demonstrate that this can lead to a large increase of the wind
  entropy, density, and temperature, and to a strong deceleration of the wind
  expansion. The consequences of this phenomenon for the possible r-process
  nucleosynthesis in the late wind still need to be explored in detail.
  Two-dimensional models show that the wind-ejecta collision is highly
  anisotropic and could lead to a directional dependence of the nucleosynthesis
  even if the neutrino-driven wind itself is spherically symmetric.

}
\FullConference{International Symposium on Nuclear Astrophysics - Nuclei in the Cosmos - IX\\
                 25-30 June 2006\\
                 CERN}

\begin{document}

\section{Introduction}
The site or sites where the r-process nucleosynthesis takes place are still
unclear. To be a major source of r-process elements, an environment must
fulfill two requirements: Firstly, it has to reproduce the r-process abundances
pattern, and secondly, it should yield the total amount of r-process material
in the present Galaxy. Due to this second constraint, several scenarios (e.g.
neutron star mergers, GRB, disks) are not very likely to be the main site,
although r-process elements are possibly produced there
\cite{Qian00,Freiburghaus99,Cameron01}. On the other hand, models for the
core collapse supernova scenario, which was proposed already in 1957 by
\cite{B2FH} and \cite{Cameron57} as the r-process site, have so far not
achieved to produce extreme conditions needed to generate the heaviest elements
observed.

In principle, the supersonic neutrino-driven wind, which forms after the onset
of the explosion, provides favorable conditions for the r-process -- a high
neutron abundance, short dynamical time scales, and high entropies. Several
groups have studied this site by means of hydrodynamic supernova explosion
simulation \cite{Woosley94,Takahashi94}, or by solving the stationary wind
equations \cite{Otsuki00,Thompson01}.

In the past years, parametric models based on the solution of the
general-relativistic steady-state wind equations have been developed
\cite{Otsuki00,Thompson01}. The relativistic treatment resulted in somewhat
higher entropies than those found in Newtonian calculations \cite{Otsuki00}.
% They assumed an inner boundary at the neutron star surface and a
% rather simplified neutrino transport, but on the other hand they had general
% relativity equations, what has been shown to have a big influence in increasing
% the wind entropies compared to the Newtonian treatment (Otsuki). 
However, also with this effect the common problem of all these studies remains
that for ``realistic parameters'' (e.g. typical neutron star masses and radii)
they do not find entropies sufficiently high for the r-process. However, none
of these studies treated possible effects due to the interaction of the wind
with preceding post-shock ejecta consistently. This interaction leads to the
formation of a wind termination shock (or reverse shock) \cite{Janka96,Tomas04},
which can change the neutrino wind properties considerably.

% What is still missing in the wind models is the possible effect due
% to the interaction of the wind with post-shock ejecta.

% The possible consequences of a wind termination
% for the nucleosynthesis has been already 
% pointed out by Qian \& Woosley, Janka \& M\"uller, Thompson et al. Even some
% work was done in this direction by means of an artificial outer boundary
% condition (Terasawa, Otsuki).

We have performed one- and two-dimensional hydrodynamic simulations that start
a few milliseconds after bounce and follow the evolution for several seconds
after the onset of the explosion. This allows us to study systematically the
influence of this reverse shock, which turns out to increase the wind entropies
significantly. Here we describe briefly the numerics we are using and some of
our first results.

\section{Numerical setup}

Our computational approach is described in detail in \cite{Scheck06a}. The
initial data at $\sim 10$ms after bounce are provided by Boltzmann simulations.
We replace the inner part of the neutron star ($\rho \gtrsim 10^{13}\;
\mathrm{g}/\mathrm{cm}^{3}$) by an inner boundary, which allows us to avoid
strong time step limitations and provides a degree of freedom for systematic
variations. The contraction of the inner boundary can be parametrized by a time
scale and a final neutron star radius. This can be justified by the fact that
the neutron star evolution depends on the still incompletely known dense matter
equation of state.  The core neutrino luminosity is put in ``by hand'' at the
inner boundary (below the neutrinosphere).  Its value is chosen such that the
explosion energy reaches typical values known from observations.  We use a
simplified neutrino transport treatment (a grey, characteristics-based scheme),
which reproduces the results of Boltzmann transport simulations qualitatively.
However, quantities like the electron fraction should be taken with caution.
Our hydrodynamics code is Newtonian, but we account for relativistic effects by
using corrections in the gravitational potential \cite{Marek06}. This
approximation has been shown to yield results very similar to the full general
relativistic treatment.

\section{Results}

\begin{figure*}[!ht]
  \centering
  \begin{tabular}{c}
    \includegraphics[width=10cm]{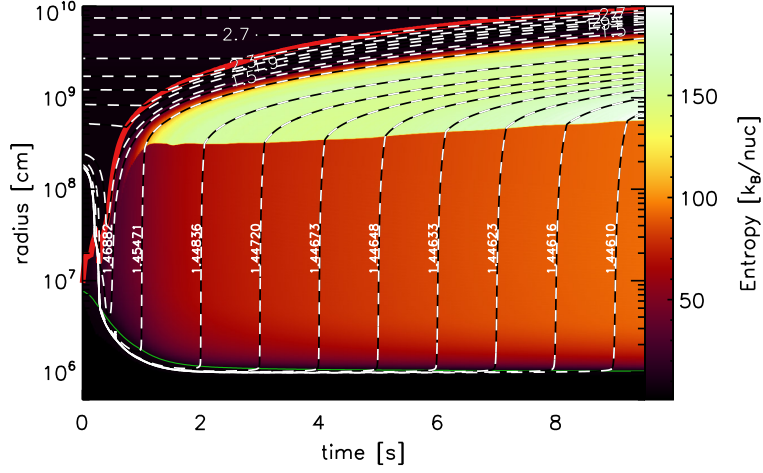}
    \end{tabular}
  \caption{Mass shell and entropy (color) plot for a one-dimensional
    model (based on a $15M_{\odot}$ progenitor) that attains an explosion
    energy of $\sim 1.2$ bethe. The green line corresponds to the neutron star
    which shrinks to a radius of $\sim 10$ km. The red line is the shock
    radius. The dashed lines are mass shell trajectories (the baryonic mass
    below a mass shell is given in $M_{\odot}$). In the wind phase the mass
    shells move to larger radii in a very short time corresponding to high wind
    velocities.  When the wind velocity becomes supersonic in the frame of the
    slower-moving material outside (i.e. in the region where the mass shell
    lines expand more slowly), a reverse shock forms.  This leads to a jump in the
    entropy (at $\sim 5\cdot10^{8}$ cm).}
  \label{fig:mass shell}
\end{figure*}

After the explosion has been launched successfully, the supernova shock
propagates outwards and the density around the neutron star decreases. Ongoing
neutrino-energy deposition leads to the formation of a neutrino-driven wind
that emerges from the neutron star surface. The matter in the wind is
accelerated to supersonic velocities, hits the slower moving preceding ejecta
and is strongly decelerated in a wind termination shock (see
Fig.\ref{fig:mass shell}).
% One can
% distinguish between wind and shocked wind, we discuss here both phases in one
% dimension and briefly also in two dimensions.

\begin{figure*}[!ht]
   \centering
   \begin{tabular}{cccc}
     \includegraphics[width=5cm]{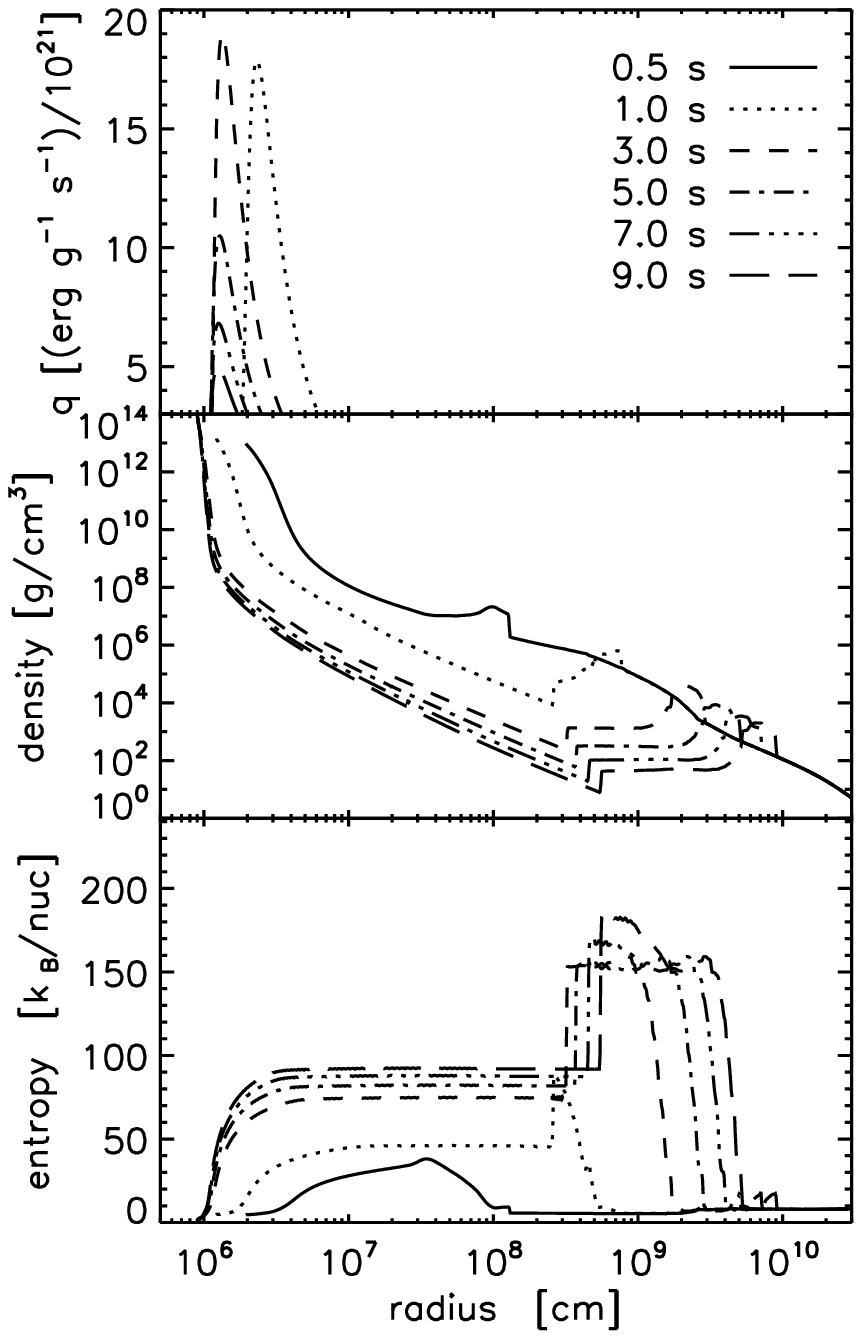}   &
     \includegraphics[width=5cm]{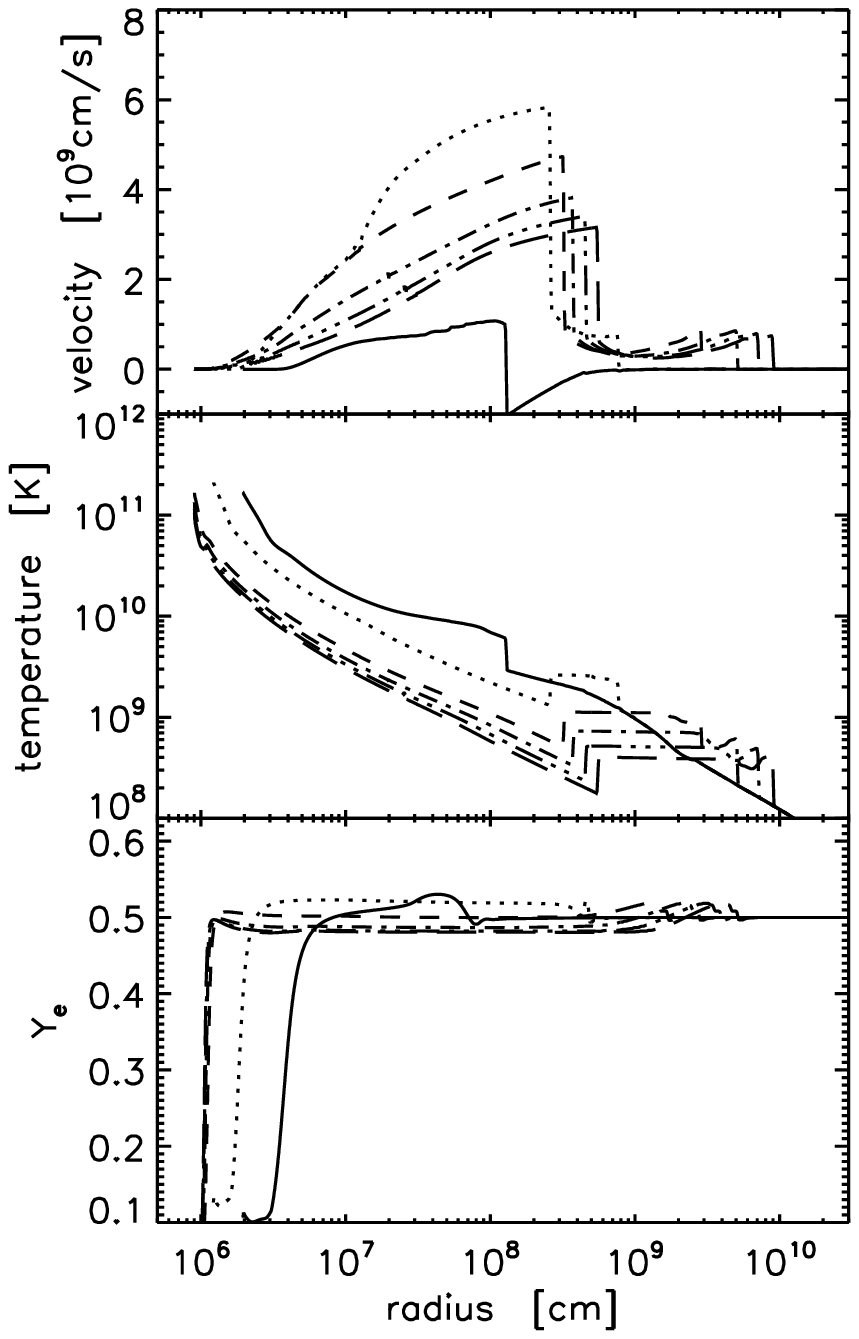}    
     \end{tabular}
   \caption{Radial profiles at different times after bounce for the neutrino
     energy deposition rate ($q$), density, entropy (left column), velocity,
     temperature and $Y_{\mathrm{e}}$ (right column). The effect of the reverse
     shock is clearly visible as a jump of the wind entropy by a factor of
     two.}
   \label{fig:profiles}
\end{figure*}

Figure \ref{fig:profiles} shows characteristic quantities from a
one-dimensional simulation as functions of radius for different times. In the
\emph{wind phase}, the velocity increases, becomes supersonic, and tends to an
asymptotic value. Also the entropy reaches a constant maximum value where the
neutrino energy deposition rate ($q$ in Fig.\ref{fig:profiles}) becomes
negligible. The quantities characteristic for the wind (velocity, entropy, ...)
are qualitatively and, for the same values for the neutrino heating rate, even
quantitatively in agreement with general relativistic stationary wind solutions
(see e.g. \cite{Thompson01}). The wind entropy increases with time due to the
decrease of the neutrino luminosity (see \cite{Qian96} for an analytic
discussion of how the entropy depends on the neutrino properties). The highest
wind entropy obtained in our simulations is not as high as those found in
previous wind studies. The reason is that the neutrino luminosities even at the
end of our simulations are higher than the lowest values considered for
stationary wind solutions.  This explains also the larger mass flux in the wind
($\dot{M}\approx 10^{-4} M_{\odot}/\mathrm{s}$) in our case.  Another
nucleosynthesis-relevant quantity, the electron fraction $Y_{\mathrm{e}}$,
evolves from proton richness ($Y_{\mathrm{e}}>0.5$) just after the onset of the
explosion (as found also in Boltzmann simulations with artificial explosions
\cite{Buras06a,Froehlich06}) to neutron richness ($Y_{\mathrm{e}}<0.5$) at $t >
3$s after bounce.  In the simulation presented here we have short expansion
time scales which should prevent a large fraction of alpha particles from
recombination to seed nuclei.  Consequently the situation favors many free
neutrons to remain unbound and thus a high neutron-to-seed ratio after the
$\alpha$-rich freeze-out phase.

% The main difference in our radial profiles compared to standard wind solutions
% is the \emph{wind termination shock}. The existence of this termination is not
% new, it was already addressed and found in previous works: Janka \& M\"uller
% found this reverse shock, Qian \& Woosly checked the effect of an outer
% boundary in their analytic studies, Thomas et al. studied the influence of the
% reverse shock on the neutrino signal generated by supernovae, and Terasawa
% analysed the influence of a wind termination by means of a stationary wind
% model.  Therefore our results are new only in the sense that we perform
% hydrodynamic simulation starting at a few milliseconds after bounce and follow
% the long-term evolution, instead of just looking for stationary wind solutions.
% Moreover, the resolution has been improved since the simulations of Janka \&
% M\"uller.  These two facts allow us to follow the reverse shock evolution.

In Figure \ref{fig:profiles} one can also see how the supersonic outflow hits
the slower preceding supernova ejecta, and is decelerated in the wind
termination shock, which leads to a sudden increase of density, temperature,
and entropy. From the Rankine-Hugoniot conditions it is straight forward to
demonstrate that the jump in the entropy increases with the square of the wind
velocity. Temperature and density decay with time much more slowly after the
wind matter was decelerated by the reverse shock. The conditions established
during that  phase ($T_{9} \sim 0.5-1$, $\rho \sim 100-10^{4}
\mathrm{g}/\mathrm{cm}^{3}$) suggest possibly important consequences for the
nucleosynthesis in the supernova ejecta. Detailed nucleosynthesis calculations,
however, are needed to explore the exact effects.

% For the r-process to take place, it is not only essential that the entropy
% reaches high values, but also the temperature is relevant.  Figure
% \ref{fig:profiles} shows that the temperature is almost constant and around
% $10^{9}$ K (below the alpha freeze out temperature) in the high-entropy region
% outside of the reverse shock. These conditions seem to be favorable for the
% r-process.

We have performed simulations of the kind described here, with varied
conditions at the inner boundary (different neutron star contractions and
neutrino luminosities) for different explosion energies and several
progenitor stars. In all cases the wind termination shock developed and had
qualitatively similar consequences. Interesting quantitative differences will
be analyzed in a paper in preparation.

% We performed simulations like the one discussed above for different neutron
% star contractions, explosion energies, and progenitor stars. In all of these
% simulations a reverse shock formed. The effect of the reverse shock is always
% important but depends on the model parameters. These dependences are analyzed
% in detail in a paper in preparation.

\begin{figure*}[!ht]
  \centering
  \begin{tabular}{cc}
    \includegraphics[width=6.5cm]{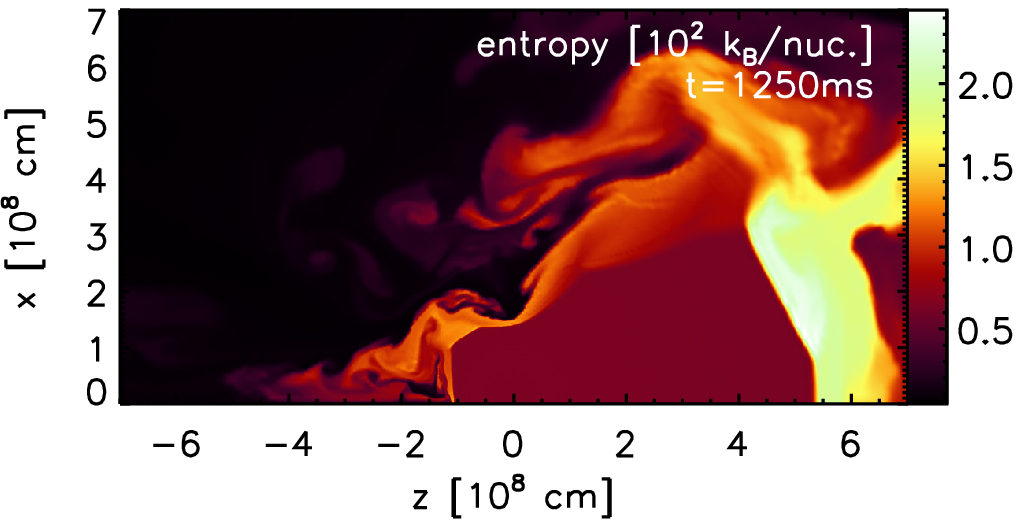}   &
    \includegraphics[width=6.5cm]{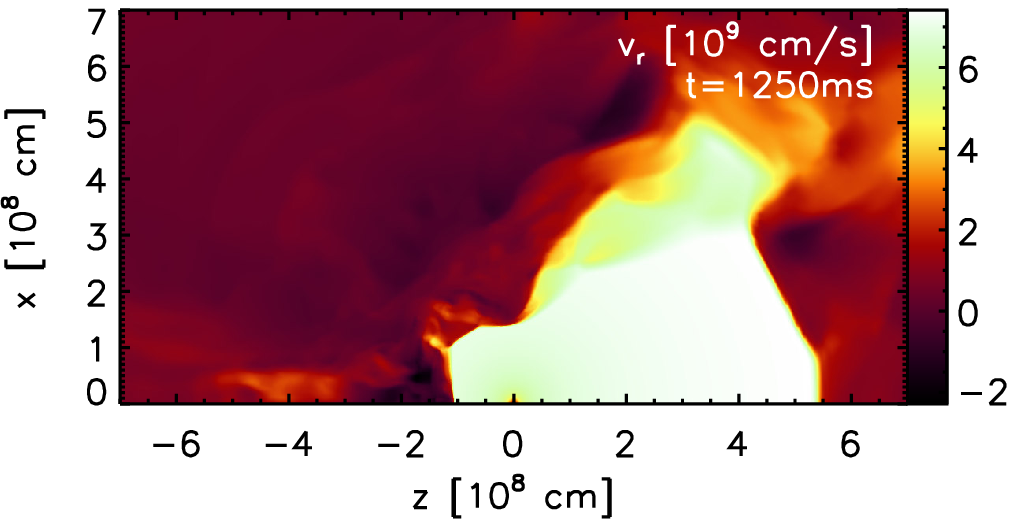}   
  \end{tabular}
  \caption{Two dimensional simulations. Entropy and velocity  at 1.2~
    seconds after core bounce. Note the highly aspherical distribution of the
    ejecta. The effect of the reverse shock on the entropy leads to similar
    maximum entropies as in one dimension.}
  \label{fig:2d}
\end{figure*}

The reverse shock forms also in our \emph{two-dimensional simulations} (see
Fig.~\ref{fig:2d} for an example). Interestingly, the conditions are strongly
dependent on the direction and suggest a very anisotropic distribution of the
nucleosynthetic products of the neutrino-driven winds. This also raises the
question how robust the condition can be in this environment for r-process
nucleosynthesis.

% parts of it are oriented at
%  an oblique angle to the (radial) wind, which leads to a weaker shock and
%  therefore a smaller entropy jump. Also the angular variation of the reverse
%  shock radius modifies the entropy distribution, compared to the one-dimensional
%  case.

\section{Conclusions}

Our work demonstrates that over a wide variation of conditions (neutron star
parameters, progenitors, spherical symmetry or multi-dimensional environment)
the wind termination shock may have non-negligible influence on the
nucleosynthesis conditions in the neutrino-driven winds. It does not only
increase the wind entropy by up to a factor of two, but also changes the time
evolution of density and temperature in an interesting way. Nucleosynthesis
calculations are needed to study the exact consequences of this so far
incompletely explored feature of supernova ejecta dynamics.

% The wind phase is reproducible by our hydrodynamic simulation, being in agreement
% with previous works. We have found that the wind termination or reverse shock
% has a crucial influence in increasing the wind entropy by a factor of two. But
% this is not the final answer since the transport treatment is still simplify
% (electron fraction can be different from what we got) and we do not consider
% the neutron star. Therefore, we have to wait till Boltzmann transport
% simulations give exact results for the wind phase. Nevertheless, it is clear
% the big influence of the reverse shock and, until detailed transport simulation
% can follow the explosion for a few seconds, the effect of a wind termination
% should be considered in future works.

% Network calculation for our mass trajectories have to be done to confirm or to
% discard the importance of the reverse shock for the r-process production. Also
% more simulations for a wide range of progenitors are required. 

{\footnotesize
\acknowledgments
Support by the Sonderforschungsbereich 375 on "Astroparticle Physics" of the
Deutsche Forschungsgemeinschaft is acknowledged. The computations were
performed on the IBM p690 clusters of the Rechenzentrum Garching and the
John-von-Neumann Institute for Computing in J\"ulich.
}

\def\aj{AJ}%
          % Astronomical Journal
\def\araa{ARA\&A}%
          % Annual Review of Astron and Astrophys
\def\apj{ApJ}%
          % Astrophysical Journal
\def\apjl{ApJ}%
          % Astrophysical Journal, Letters
\def\apjs{ApJS}%
          % Astrophysical Journal, Supplement
\def\ao{Appl.~Opt.}%
          % Applied Optics
\def\apss{Ap\&SS}%
          % Astrophysics and Space Science
\def\aap{A\&A}%
          % Astronomy and Astrophysics
\def\aapr{A\&A~Rev.}%
          % Astronomy and Astrophysics Reviews
\def\aaps{A\&AS}%
          % Astronomy and Astrophysics, Supplement
\def\azh{AZh}%
          % Astronomicheskii Zhurnal
\def\baas{BAAS}%
          % Bulletin of the AAS
\def\jrasc{JRASC}%
          % Journal of the RAS of Canada
\def\memras{MmRAS}%
          % Memoirs of the RAS
\def\mnras{MNRAS}%
          % Monthly Notices of the RAS
\def\pra{Phys.~Rev.~A}%
          % Physical Review A: General Physics
\def\prb{Phys.~Rev.~B}%
          % Physical Review B: Solid State
\def\prc{Phys.~Rev.~C}%
          % Physical Review C
\def\prd{Phys.~Rev.~D}%
          % Physical Review D
\def\pre{Phys.~Rev.~E}%
          % Physical Review E
\def\prl{Phys.~Rev.~Lett.}%
          % Physical Review Letters
\def\pasp{PASP}%
          % Publications of the ASP
\def\pasj{PASJ}%
          % Publications of the ASJ
\def\qjras{QJRAS}%
          % Quarterly Journal of the RAS
\def\skytel{S\&T}%
          % Sky and Telescope
\def\solphys{Sol.~Phys.}%
          % Solar Physics
\def\sovast{Soviet~Ast.}%
          % Soviet Astronomy
\def\ssr{Space~Sci.~Rev.}%
          % Space Science Reviews
\def\zap{ZAp}%
          % Zeitschrift fuer Astrophysik
\def\nat{Nature}%
          % Nature
\def\iaucirc{IAU~Circ.}%
          % IAU Cirulars
\def\aplett{Astrophys.~Lett.}%
          % Astrophysics Letters
\def\apspr{Astrophys.~Space~Phys.~Res.}%
          % Astrophysics Space Physics Research
\def\bain{Bull.~Astron.~Inst.~Netherlands}%
          % Bulletin Astronomical Institute of the Netherlands
\def\fcp{Fund.~Cosmic~Phys.}%
          % Fundamental Cosmic Physics
\def\gca{Geochim.~Cosmochim.~Acta}%
          % Geochimica Cosmochimica Acta
\def\grl{Geophys.~Res.~Lett.}%
          % Geophysics Research Letters
\def\jcp{J.~Chem.~Phys.}%
          % Journal of Chemical Physics
\def\jgr{J.~Geophys.~Res.}%
          % Journal of Geophysics Research
\def\jqsrt{J.~Quant.~Spec.~Radiat.~Transf.}%
          % Journal of Quantitiative Spectroscopy and Radiative Trasfer
\def\memsai{Mem.~Soc.~Astron.~Italiana}%
          % Mem. Societa Astronomica Italiana
\def\nphysa{Nucl.~Phys.~A}%
          % Nuclear Physics A
\def\physrep{Phys.~Rep.}%
          % Physics Reports
\def\physscr{Phys.~Scr}%
          % Physica Scripta
\def\planss{Planet.~Space~Sci.}%
          % Planetary Space Science
\def\procspie{Proc.~SPIE}%
          % Proceedings of the SPIE
\let\astap=\aap
\let\apjlett=\apjl
\let\apjsupp=\apjs
\let\applopt=\ao

%\begin{thebibliography}{10}
\bibliographystyle{unsrt}
\bibliography{biblio}

\end{document}